\documentstyle[12pt]{article}
\makeatletter
\def\fmslash{\@ifnextchar[{\fmsl@sh}{\fmsl@sh[0mu]}}
\def\fmsl@sh[#1]#2{%
  \mathchoice
    {\@fmsl@sh\displaystyle{#1}{#2}}%
    {\@fmsl@sh\textstyle{#1}{#2}}%
    {\@fmsl@sh\scriptstyle{#1}{#2}}%
    {\@fmsl@sh\scriptscriptstyle{#1}{#2}}}
\def\@fmsl@sh#1#2#3{\m@th\ooalign{$\hfil#1\mkern#2/\hfil$\crcr$#1#3$}}
\makeatother
\global\arraycolsep=2pt 
\def\dlr{\stackrel{\leftrightarrow}{D}}
\begin{document}
\thispagestyle{empty}
\vspace*{-1.9cm}
\begin{flushright}
{\small TTP 96-27 \\
hep-ph/9607299}
\end{flushright}

\begin{center}
{\bf Effective Theory for Heavy Quarkonia Decays}
\vspace*{1cm} \\
{\sc Thomas Mannel} \\
{\it Institut f\"ur Theoretische Teilchenphysik, Universit\"at
Karlsruhe\\
D--76128 Karlsruhe, Germany}
\vspace*{1cm} \\
Contribution to the {\it III German--Russian 
Workshop on Heavy Quark Physics}, \\ Dubna, 
Russia, May 20--22, 1996
\end{center}
\vspace*{0.85cm}
\begin{abstract}
{\small  An effective theory approach 
         to heavy quarkonia decays based on the $1/m_Q$ expansion 
         is introduced. Its application to decays in which the two heavy 
         quarks annihilate is discussed.}
\end{abstract}

\vspace*{0.25cm}
\section{Introduction}
Heavy Quark Effective Theory (HQET) \cite{HQET} has turned out to be a very 
successful approach to describe systems with a single heavy quark. 
It is based on the infinite mass limit of QCD which serves as 
a starting point to perform a systematic expansion in 
$\Lambda_{QCD} / m_Q$ and $\alpha_s (m_Q)$ using the methods of 
effective field theory. 

Presently we are at the threshold of the discovery of flavoured 
``doubly heavy'' systems, i.e.\ states consisting of two heavy quarks
(such as the $B_c$'s and also baryons with $b=c=\pm 1$, $b=\pm 2$ or $c=\pm 2$), 
while quarkonia like systems (the $\psi$'s and the $\Upsilon$'s) 
are in the meantime quite well known. 

Motivated by these expectations the question arises whether one may 
set up a similar effective theory approach for systems with two 
(or even more) heavy quarks based on the $1/m_Q$ expansion of QCD.     
It turns out \cite{KMO} that one can not use the static limit for
two heavy quarks if their velocities differ only by an amount
of order $1/m_Q$, i.e. $vv' - 1 \sim \Lambda_{QCD} / m_Q$. 
The static limit breaks down and one is forced to include at 
least the kinetic energy into the leading order dynamics; in other 
words one has to use a non-relativistic approximation 
instead of the static limit. Such approximations have been formulated 
in the QED context (NRQED) some time ago; Two slightly 
different approaches (NRQCD \cite{BBL} and HQQET \cite{MS})
have recently been studied in QCD. 
In this talk the general features of such an effective
theory is outlined and its application to this class of decays is 
considered in which the two heavy quarks annihilate.

\section{Structure of HQQET/NRQCD}

In order to set up a heavy quarkonium effective theory one starts from 
the $1/m_Q$ expansion of the QCD Lagrangian and the corresponding 
expansion of the fields
\begin{eqnarray} \label{FWfield}
&& Q(x) =  e^{-im_Qv\cdot x}\left[ 1 +\frac{1}{2m_Q}(i\fmslash{D}_{\perp}) +
\frac{1}{4m_Q^2}\left( v\cdot D \fmslash{D}_{\perp} - \frac{1}{2}
\fmslash{D}_{\perp}^2 \right) + \cdots \right] h_v (x)\,,
\\ 
&& {\cal L} = \bar h_v\left[ iv \cdot D - \frac{1}{2m_Q}
\fmslash{D}_{\perp}^2
 + \frac{i}{4m_Q^2}\left(-\frac{1}{2}
\fmslash{D}_{\perp}^2 v \cdot D +
\fmslash{D}_{\perp} v\cdot D 
\fmslash{D}_{\perp} - \frac{1}{2} v\cdot D 
\fmslash{D}_{\perp}^2 \right) + \cdots \right] h_v\,.\nonumber
\end{eqnarray}
The first terms of these two expansions define the static limit, 
which has been successfully applied to systems with a single heavy 
quark. In order to describe a system with more than one heavy (anti)quark
one has to write down the same expansion (\ref{FWfield}) for each heavy 
quark. However, in the static limit for a state with two or more 
heavy quarks one runs into problems with diverging phases and ``complex 
anomalous dimensions'', which are considered in detail in \cite{KMO}. 

In order to cure this problem one has to choose the unperturbed 
system such that these phases are already generated by the leading order
dynamics, i.e.\ instead of the static limit one has to use the 
non-relativistic Lagrangian. For a system with a heavy quark and a 
heavy antiquark one then starts from 
\begin{equation}
{\cal L}_0   = \bar{h}_v^{(+)} (ivD) h^{(+)}_v - \bar{h}_v^{(-)} (ivD) h^{(-)}_v
              + K_1 \, , \quad 
K_1 = \bar{h}_v^{(+)} \frac{(iD)^2}{2m_Q} h^{(+)}_v
             + \bar{h}_v^{(-)} \frac{(iD)^2}{2m_Q} h^{(-)}_v
\label{lnull}
\end{equation}
where we have assumed for simplicity that the two quarks have the 
same mass; the case of unequal mass is obvious.

Most of the success of HQET is due to heavy quark flavour and spin symmetry.
However, once one uses (\ref{lnull}) the symmetries are somewhat different
for HQQET. First of 
all, (\ref{lnull}) depends on the mass through the kinetic energy 
term; consequently 
the states will depend on $m_Q$ in a non-perturbative
way and heavy flavour symmetry is lost. On the other hand, 
(\ref{lnull}) does not depend on the spins of the two heavy quarks 
so there is a spin symmetry which is larger than in HQET because we 
have two heavy quark spins; the resulting symmetry is an 
$SU(2) \otimes SU(2)$ corresponding to separate rotations of the 
two spins. 

For the case of heavy quarkonia all states fall into spin symmetry
quartets which should be degenerate in the non-relativistic limit. 
In spectroscopic notation  ${}^{2S+1}\ell_J$
these quartets consist of the states
\begin{equation}
[n{}^1 \ell_\ell \quad n{}^3 \ell_{\ell-1} \quad n{}^3 \ell_\ell \quad
 n{}^3 \ell_{\ell+1} ]
\label{fourstates}
\end{equation}
For the ground states the spin symmetry quartet consists of the $\eta_Q$ 
(the $0^-$ state) and the three polarization directions of the 
$\Upsilon_Q$ (the $1^-$ state). 

The heavy quarkonia spin symmetry restricts the non-perturbative 
input to a calculation of processes involving heavy quarkonia. 
Of particular interest are decays in which the heavy quarks inside
the heavy quarkonium annihilate. The annihilation is a short 
distance process that can be calculated perturbatively in terms of 
quarks and gluons, while the long distance contribution is encoded in 
certain matrix elements of quark operators. Logarithmic dependences
on the heavy quark mass may be calculated by employing the usual 
renormalization group machinery.   

\section{Annihilation Decays of Heavy Quarkonia}

The starting point to calculate processes like 
$\eta_Q \to $ {\it light hadrons}, 
$\eta_Q \to \gamma \, + $ {\it light hadrons}, 
or the corresponding decays of the $\Upsilon_Q$ states
is the transition operator $T$ for two heavy quarks 
which annihilate into light degrees of freedom. This will in general 
be bilinear in the heavy quark fields, such that   
\begin{equation} \label{Top}
T (X,\xi) = (-i) \bar{Q}(X+\xi) K (X,\xi)  Q(X-\xi) 
\end{equation}
where $K (X,\xi)$ involves only light degrees of freedom and $X$ and $\xi$
correspond to the cms and relative coordinate respectively. We identify 
the field $Q$ with the quark and $\bar{Q}$ with the antiquark, so 
we shall make the large scale $m_Q$ explicit by redefining the fields
as 
\begin{equation}
Q (x) = \exp(-im_Q (vx) ) Q_v^{(+)} (x) , \quad
\bar{Q} (x) =  \exp(-im_Q (vx) ) \bar{Q}_{v}^{(-)} (x)
\end{equation}
This corresponds to the usual splitting of the heavy quark momentum 
into a large part $m_Q v$ and a residual piece $k$. Inserting this into 
(\ref{Top}) this yields
\begin{equation}
 T = (-i)  \exp[-i2 m_Q v X] 
     \bar{Q}_v^{(-)} (X+\xi) K (X,\xi)  Q_v^{(+)} (X-\xi)
\end{equation}
The inclusive decay rate for the decay of a quarkonium 
$\Psi \to$ {\it light degrees of freedom} is then given by
\begin {equation}
\Gamma = \langle \Psi | 
\left[ \int d^4 X d^4 \xi d^4 \xi ' T (X,\xi) T^\dagger (0,\xi ') 
         + {\rm h.c.} \right] | \Psi \rangle
\end{equation}    
The next step is to perform 
an Operator Product Expansion (OPE) for the non-local product of the 
quark field operators. This expansion will yield four-quark operators 
of increasing dimension starting with dim-6 operators. The increasing 
dimension of these operators will be compensated by inverse powers of 
the heavy quark mass, so generically the rate takes the form
\begin {equation}
\Gamma = m_Q \sum_{n,i} \left(\frac{1}{m_Q} \right)^{n-2} 
         C ({\cal O}_i^{(n)},\mu) \langle \Psi | 
         {\cal O}_i^{(n)}| \Psi \rangle|_\mu
\end{equation}   
where $n=6,7,\cdots$ is the dimension of the operator and $i$ labels 
different operators with the same dimension. The coefficients  
$C ({\cal O}_i^{(n)},\mu)$ are related to the 
short distance annihilation process 
and hence may be calculated in perturbation theory in terms of quarks
and gluons. Once QCD radiative corrections are included, the 
$C ({\cal O}_i^{(n)},\mu)$ acquire a dependence on 
the renormalization scale 
$\mu$ which is governed by the renormalization group of the effective 
theory. The rate $\Gamma$ is independent of $\mu$ and hence the
$\mu$ dependence of $C ({\cal O}_i^{(n)},\mu)$ has to be compensated by 
a corresponding dependence of the matrix elements.     

The non-perturbative contributions are encoded in the 
matrix elements of the local four-quark operators, and the mass dependence 
of these operators is also expanded in powers of $1/m_Q$ and thus  the 
remaining $m_Q$ dependence of the matrix elements is only due to the 
states. In terms of the $m_Q$ independent static fields 
$$
Q_v^{(+)} (x) = h^{(+)} (x) + {\cal O} (1/m_Q) \quad
Q_v^{(-)} (x) = h^{(-)} (x) + {\cal O} (1/m_Q)
$$
one has in total four dim-6 operators
$$
A_1 ^{(C)} = \bar{h}^{(+)} \gamma_5 C h^{(-)} \, 
               \bar{h}^{(-)} \gamma_5 C h^{(+)} \mbox{ and }
A_2 ^{(C)} = \bar{h}^{(+)} \gamma_\mu C h^{(-)} \,
               \bar{h}^{(-)} \gamma^\mu C h^{(+)} 
$$
where $C$ is a color matrix, where one has the two possibilities
$C \otimes C = 1 \otimes 1$ or $C \otimes C = T^a \otimes T^a$.
These operators do not mix under renormalization, all anomalous dimensions
vanish. 

There are no dim-7 operators, since these are all proportional to 
$(ivD)$ and can be rewritten in terms of dim-8 operators by the equations 
of motion. At dim-8 one finds 30 local operators
\begin{eqnarray*}
B_1^{(C)} &=& [ (iD_\mu) \bar{h}^{(+)} \gamma_5 C h^{(-)}] \, 
              [ (iD^\mu) \bar{h}^{(-)} \gamma_5 C h^{(+)}] \\
B_2^{(C)} &=& [ (iD_\mu) \bar{h}^{(+)} \gamma_\lambda C h^{(-)}] \, 
              [ (iD^\mu) \bar{h}^{(-)} \gamma^\lambda C h^{(+)}] \\
B_3^{(C)} &=& [ (iD_\mu) \bar{h}^{(+)} \gamma_\lambda C h^{(-)}] \, 
              [ (iD^\lambda) \bar{h}^{(-)} \gamma^\mu C h^{(+)}] \\
 \nonumber \\
C_1^{(C)} &=& [ (iD^\mu) \bar{h}^{(+)} \gamma_5 C h^{(-)}] \, 
              [ \bar{h}^{(-)} \gamma_5 C (i\dlr_\mu) h^{(+)}] + {\rm h.c.} \\
C_2^{(C)} &=& [ (iD^\mu) \bar{h}^{(+)} \gamma_\mu C h^{(-)}] \, 
              [ \bar{h}^{(-)} C (i\fmslash{\dlr}) h^{(+)}] + {\rm h.c.} \\
C_3^{(C)} &=& [ (iD^\lambda) \bar{h}^{(+)} \gamma^\mu C h^{(-)}] \, 
              [ \bar{h}^{(-)} \gamma_\mu C (i\dlr)_\lambda h^{(+)}] 
              + {\rm h.c.} \\
C_4^{(C)} &=& [ (iD^\lambda) \bar{h}^{(+)} \gamma^\mu C h^{(-)}] \, 
              [ \bar{h}^{(-)} \gamma_\lambda C (i\dlr)_\mu h^{(+)}] 
              + {\rm h.c.} \\
 \nonumber \\
D_1^{(C)} &=& [ \bar{h}^{(+)} \gamma_5 C (i\dlr)_\mu h^{(-)}] \, 
              [ \bar{h}^{(-)} \gamma_5 C (i\dlr)^\mu h^{(+)}] \\
D_2^{(C)} &=& [ \bar{h}^{(+)} C (i\fmslash{\dlr}) h^{(-)}] \, 
              [ \bar{h}^{(-)} C (i\fmslash{\dlr}) h^{(+)}] \\
D_3^{(C)} &=& [ \bar{h}^{(+)} \gamma_\lambda C (i\dlr)_\mu h^{(-)}] \, 
              [ \bar{h}^{(-)} \gamma^\lambda C (i\dlr)^\mu h^{(+)}] \\
D_4^{(C)} &=& [ \bar{h}^{(+)} \gamma_\mu C (i\dlr)_\lambda h^{(-)}] \, 
              [ \bar{h}^{(-)} \gamma^\lambda C (i\dlr)^\mu h^{(+)}] \\
 \nonumber \\
E_1^{(C)} &=& [ \bar{h}^{(+)} \gamma_5 C h^{(-)}] \, 
              [ \bar{h}^{(-)} \gamma_5 C (i\dlr)^2 h^{(+)}] + {\rm h.c.} \\
E_2^{(C)} &=& [ \bar{h}^{(+)} \gamma^\mu C h^{(-)}] \, 
              [ \bar{h}^{(-)} C (i\fmslash{\dlr}) (i\dlr)_\mu h^{(+)}]
              + {\rm h.c.} \\
E_3^{(C)} &=& [ \bar{h}^{(+)} \gamma^\mu C h^{(-)}] \, 
              [ \bar{h}^{(-)} C (i\dlr)_\mu (i\fmslash{\dlr}) h^{(+)}]
              + {\rm h.c.} \\
E_4^{(C)} &=& [ \bar{h}^{(+)} \gamma^\mu C h^{(-)}] \, 
              [ \bar{h}^{(-)} \gamma_\mu C (i\dlr)^2 h^{(+)}] + {\rm h.c.} 
\end{eqnarray*}
In addition to these contributions one has also non-local terms originating
from single insertions of the Lagrangian of order $1/m_Q^2$ and from 
double insertions of the Lagrangian of order $1/m_Q$. Under 
renormalization the local dim-8 operators do not mix; only the double 
insertion of the kinetic energy operator of order $1/m_Q$ mixes into 
some of the above operators. Denoting this contribution as $T^{(C)}_i$
\begin{equation}
T^{(C)}_i = \frac{(-i)^2}{2} \int d^4 x d^4 y T[A_i^{(C)} K_1(x) K_1 (y)]
\end{equation}
one obtains in one-loop renormalization group improved perturbation theory
two sets of equations for the coefficients of the operators with the 
spin structure $\gamma_5 \otimes \gamma_5$  
\begin{eqnarray*}
C(D_1^{(1)},\mu) &=& C(D_1^{(1)},m_Q) + \frac{32}{9} 
                     \frac{1}{33-2n_f} C(T_1^{(8)},m_Q) \ln \eta \\
C(E_1^{(1)},\mu) &=& C(E_1^{(1)},m_Q) - 
                     \frac{8}{33-2n_f} C(T_1^{(1)},m_Q) \ln \eta \\
C(B_1^{(8)},\mu) &=& C(B_1^{(8)},m_Q) - 
                     \frac{24}{33-2n_f} C(T_1^{(8)},m_Q) \ln \eta \\
C(D_1^{(8)},\mu) &=& C(D_1^{(8)},m_Q) - 
                     \frac{16}{33-2n_f} C(T_1^{(1)},m_Q) \ln \eta +
                 \frac{20}{3}\frac{1}{33-2n_f} C(T_1^{(1)},m_Q) \ln \eta \\
C(E_1^{(8)},\mu) &=& C(E_1^{(8)},m_Q) - \frac{14}{3}
                     \frac{1}{33-2n_f} C(T_1^{(1)},m_Q) \ln \eta 
\end{eqnarray*}
where $\eta = (\alpha_s (\mu) / \alpha_s (m_Q) )$. Furthermore, the 
coefficients $C(T_i^{(C)},m_Q)$ are the same as the ones for the dim-6 
operators $C(A_i^{(C)},m_Q)$ since the kinetic energy operator is not 
renormalized.

The second set of equations is for the operators with spin structure
$\gamma_\mu \otimes \gamma^\mu$ and due to heavy quarkonia spin symmetry 
one obtains the same equations; all other renormalization
group equations are trivial. 

A calculation of an annihilation decay then involves to calculate the 
$C ({\cal O}_i^{(n)},\mu)$ at the scale $\mu = m_Q$ by 
matching the effective theory to full QCD. Once this is done, one may 
run down to some small scale $\mu$ of the order of the ``binding energy''
of the heavy quarkonium, thereby resumming the well known logarithms of the 
form $\ln(m_Q / \mu)$ that appear in the calculations of decay rates of 
heavy $p$-wave quarkonia. As an example, in \cite{MW}
the decay $\eta_Q \to \gamma$ + {\it light hadrons} is studied in HQQET.

The matrix elements of these operators are non-perturbative quantities, 
which are constrained by heavy quarkonia spin symmetry. 
In order to exploit this symmetry, one may use the 
usual representation matrices for the spin singlet and spin triplet 
quarkonia
\begin{equation}
H_1 (v) = \sqrt{M} P_+ \gamma_5 \mbox{ for } S = 0 \, , \quad
H_3 (v) = \sqrt{M} P_+ \epsilon \mbox{ for } S = 1 
\end{equation}
where $M \approx 2m_Q$ is the mass of the heavy quarkonium and 
$P_+ = (1+\fmslash{v})/2$. Using this one finds for the matrix elements 
of the dim-6 operators
\begin{equation}
\langle \Psi | \bar{h}^{(+)} \Gamma C h^{(-)}\, 
               \bar{h}^{(-)} \Gamma ' C h^{(+)}  | \Psi \rangle
= a^{(C)} (n,\ell) G \, , \mbox{ with } G =  
{\rm Tr} (\overline{H}_{2s+1} \Gamma ) 
                  {\rm Tr} (\Gamma ' H_{2s+1} ) 
\end{equation}
Thus for each $n$ and $\ell$ and for each color combination one finds 
a single parameter for both the spin singlet and spin triplet quarkonium.

Correspondingly one finds for the dim-8 operators
\begin{eqnarray*}
\langle \Psi| (iD_\mu) [\bar{h}^{(+)} \Gamma C h^{(-)}]\, 
            (iD_\nu) [\bar{h}^{(-)} \Gamma ' C h^{(+)}]  | \Psi  \rangle
&=& b^{(C)} (n,\ell) (g_{\mu \nu} - v_\mu v_\nu ) G \\
\langle \Psi| [\bar{h}^{(+)} \Gamma C (i \dlr _\mu) h^{(-)}]\, 
(iD_\nu) [\bar{h}^{(-)} \Gamma ' C h^{(+)}] + {\rm h.c.} | \Psi \rangle
&=& c^{(C)} (n,\ell) (g_{\mu \nu} - v_\mu v_\nu ) G \\
\langle \Psi | [\bar{h}^{(+)} \Gamma C (i \dlr _\mu) h^{(-)}]\, 
 [\bar{h}^{(-)} \Gamma ' C (i \dlr _\nu) h^{(+)}]  | \Psi  \rangle
&=& d^{(C)} (n,\ell) (g_{\mu \nu} - v_\mu v_\nu ) G \\
\langle \Psi| [\bar{h}^{(+)} \Gamma C (i \dlr _\mu) 
                          (i \dlr _\nu)  h^{(-)}]\, 
[\bar{h}^{(-)} \Gamma ' C h^{(+)}] + {\rm h.c.} | \Psi  \rangle
&=& e^{(C)} (n,\ell) (g_{\mu \nu} - v_\mu v_\nu ) G 
\end{eqnarray*}
For fixed values of $n$ and $\ell$ one finds that eight parameters are needed 
to describe the matrix elements of the dim-8 operators. 

These matrix elements are non-perturbative, but from vacuum insertion one 
suspects
\begin{eqnarray*} 
 a^{(1)} (n,0) \sim |R_{n0} (0) |^2 \gg  a^{(1)} (n,\ell) 
\mbox{ for } \ell \ne 0 \, ,
\quad 
 a^{(1)} (n,0) \gg  a^{(8)} (n,\ell) \mbox{ for all } n,\ell \\
d^{(1)} (n,1) \sim |R^\prime_{n1} (0) |^2 \gg d^{(1)} (n,\ell) 
\mbox{ for } \ell \ne 1 \, , \quad
d^{(1)} (n,1) \gg  d^{(8)} (n,\ell) \mbox{ for all } n, \ell \\
e^{(1)} (n,0) \sim {\rm Re} \left[R^{\prime \prime}_{n0} (0) R^*_{n0} (0)\right] 
              \gg e^{(1)} (n,\ell)
\mbox{ for } \ell \ne 0  \, ,  \quad
e^{(1)} (n,0) \gg  e^{(8)} (n,\ell) \mbox{ for all } n, \ell 
\end{eqnarray*}
where $R_{nl} (r)$ is the radial wave function of the quarkonium. The same reasoning 
yields the expectation that $b^{(C)} (n,\ell)$ and $c^{(C)} (n,\ell)$ are 
small compared to the coefficients that are non-vanishing in vacuum insertion.

\end{document}